\documentclass[10pt,conference]{IEEEtran}
\ifCLASSINFOpdf
\else
\fi

\usepackage{graphicx}
\usepackage{bm}
\usepackage{amsfonts}
\usepackage{amsmath}
\usepackage{amssymb}
\usepackage{times}
\usepackage{cite}
\usepackage{subfigure}
\usepackage{latexsym,bm,amsmath,amssymb} 
\usepackage{CJK}
\usepackage{hhline}
\usepackage{url}
\usepackage{multirow} 
\usepackage{amsmath}
\usepackage{xcolor}
\usepackage[linesnumbered,ruled,vlined]{algorithm2e}
\usepackage{epstopdf}
\usepackage{algorithmicx}

\usepackage{geometry}
\begin{document}

%
\title{On the Pulse Shaping for Delay-Doppler Communications}

\author{\IEEEauthorblockN{Shuangyang Li\IEEEauthorrefmark{1},
Weijie Yuan\IEEEauthorrefmark{2},
Zhiqiang Wei\IEEEauthorrefmark{3},
Jinhong Yuan\IEEEauthorrefmark{4},
Baoming Bai\IEEEauthorrefmark{5},
and
Giuseppe Caire\IEEEauthorrefmark{1}
}

\IEEEauthorblockA{\IEEEauthorrefmark{1} Technische Universit{\"a}t Berlin, Berlin, Germany,}
\IEEEauthorblockA{\IEEEauthorrefmark{2} Southern University of Science and Technology, Shenzhen, China}
\IEEEauthorblockA{\IEEEauthorrefmark{3} Xi'an Jiaotong University, Xi'an, China,}
\IEEEauthorblockA{\IEEEauthorrefmark{4}University of New South Wales, Sydney, Australia,}
\IEEEauthorblockA{\IEEEauthorrefmark{5}Xidian University, Xi'an, China}
\vspace{-10mm}
}


\maketitle

\begin{abstract}
In this paper, we study the pulse shaping for delay-Doppler (DD) communications. We start with constructing a basis function in the DD domain following the properties of the Zak transform. Particularly, we show that the constructed basis functions are globally quasi-periodic while locally twisted-shifted, and their significance in time and frequency domains are then revealed. We further analyze the ambiguity function of the basis function, and show that fully localized ambiguity function can be achieved by constructing the basis function using periodic signals. More importantly, we prove that time and frequency truncating such basis functions
naturally leads to approximate delay and Doppler orthogonalities, if the truncating windows are periodic within the support. Motivated by this, we propose a DD Nyquist pulse shaping scheme considering signals with periodicity. Finally, our conclusions are verified by using various strictly or approximately periodic pulses.
\vspace{-3mm}
\end{abstract}


\IEEEpeerreviewmaketitle
\section{Introduction}
Delay-Doppler (DD) communications have recently emerged as a promising new type of communication paradigm aiming to meet the challenging requirements for future wireless communications\cite{Zhiqiang_magzine,Viterbo2022DDcommunications}. A typical delay-Doppler communication scheme is the orthogonal time frequency space (OTFS) modulation proposed by Hadani \emph{et. al}~\cite{Hadani2017orthogonal}, which has demonstrated strong delay and Doppler resilience and excellent performance over various wireless channels~\cite{Zhiqiang_magzine,Viterbo2022DDcommunications}.

The DD communication scheme multiplexes information symbols in the DD domain rather than the conventional time-frequency (TF) domain, and therefore directly explores the appealing features of the DD domain wireless channels.
As a result, DD communication enjoys many promising advantages than TF domain communications, including the simplicity of channel estimation~\cite{Fair_compare_OTFS_OFDM}, potentially-full-diversity data transmission~\cite{Raviteja2019effective}, relaxed channel code design~\cite{Li2020performance}, and low complexity precoding for MIMO transmissions~\cite{LSY_THP}. Furthermore, recent literature has also demonstrated the adaptability of DD communication to radar sensing~\cite{Saeid2023beamspace}, which makes DD communications a strong candidate for future wireless networks.

Despite the aforementioned advantages, the design of DD communication waveforms have not been fully understood in the literature. In particular, pulse shaping to convert digital symbols into analog waveforms plays an important role in the implementation of any communication system.
It should be highlighted that the DD domain signal processing is governed by the Zak transform~\cite{janssen1988zak,Bolcskei1994Gabor}, as illustrated in~\cite{MohammedBITSpart1,Mohammed2023part2}. However, how to cleverly use the property of the Zak transform to better shape the DD communication signals has not been discussed in detail.

In this paper, we study the pulse shaping for DD communications. We first construct the DD domain basis functions by using atom pulses. Different from the discussion in~\cite{Viterbo2022DDcommunications}
and~\cite{mohammed2021derivation}, our construction highlights both the global quasi-periodicity in the DD domain and the local twisted-shift in the fundamental rectangle.
Then, we derive the corresponding basis functions in time and frequency domains and show that the quasi-periodicity results in a train of pulses, while the twisted-shift translates into signal tones, which suggests that the DD communication simultaneously and symmetrically multiplexes symbols in both time and frequency.
We further investigate the ambiguity function of the constructed basis functions and show that perfectly periodic signals can possess a fully localized ambiguity function. More importantly, it is proved that when the basis functions constructed by periodic atom pulses are truncated in time and frequency, approximate orthogonality naturally emerges, if the time and frequency windows for truncation are periodic within the support. As such, we propose a DD Nyquist pulse shaping scheme considering the signal periodicity. Our conclusions are finally examined numerically using various strictly or approximately periodic pulses.

\section{Preliminaries on Zak Transform}
In this section, we will present the definition of the Zak transform and some important lemmas that will be frequently used throughout this paper. 

\textbf{Definition} (\emph{Zak Transform~\cite{janssen1988zak,Bolcskei1994Gabor}}):
Let $x\left( t \right)$ be a complex time-continuous function, and let $T$ be a positive constant. Then, the Zak transform is defined by
\begin{align}
{{\cal Z}_x}\left( {\tau ,\nu } \right)\buildrel \Delta \over = \sqrt T \sum\limits_{k =  - \infty }^\infty  {x\left( {\tau  + kT} \right){e^{ - j2\pi k\nu T}}} ,
\label{Zak_Transform_def}
\end{align}
for $ -\infty   < \tau  < \infty $ and $ -\infty  < \nu  < \infty$.

\textbf{Lemma~1} (\emph{Quasi-Periodicity~\cite{Bolcskei1994Gabor}}):
The Zak transform is quasi-periodic along the delay axis with period $T$ and periodic along the Doppler axis with period ${\frac{1}{T}}$, i.e., ${{\cal Z}_x}\left( {\tau  + T,\nu } \right) = {e^{j2\pi T\nu }}{{\cal Z}_x}\left( {\tau ,\nu }\right)$, and ${{\cal Z}_x}\left( {\tau ,\nu  + \frac{1}{T}} \right) = {{\cal Z}_x}\left( {\tau ,\nu } \right)$.

\textbf{Lemma~2} (\emph{Shift Properties~\cite{Bolcskei1994Gabor}}):
Given the original signal $x\left( t \right)$, its frequency-shifted and time-delayed version ${x_1}\left( t \right) ={e^{j2\pi {\nu _0}\left( {t - {\tau _0}} \right)}}x\left( t - {\tau _0} \right)$ has the Zak transform
\begin{align}
{{\cal Z}_{{x_1}}}\left( {\tau ,\nu } \right) = {e^{j2\pi {\nu _0}\left( {\tau  - {\tau _0}} \right)}}{{\cal Z}_x}\left( {\tau  - {\tau _0},\nu  - {\nu _0}} \right)\label{Zak_shift_time_freq}.
\end{align}

\textbf{Lemma~3} (\emph{Connections between products of Zak transform and ambiguity function~\cite{Bolcskei1994Gabor}}):
The cross ambiguity function for functions $x\left( t \right)$ and $y\left( t \right)$ is defined by
\begin{align}
{A_{x,y}}\left( {\tau ,\nu } \right) \buildrel \Delta \over = &\int_{ - \infty }^\infty  {x\left( t \right){y^*}\left( {t - \tau } \right){e^{ - j2\pi \nu \left( {t - \tau } \right)}}{\rm{d}}t}  \notag\\
=& \int_{ - \infty }^\infty  {X\left( f \right){Y^*}\left( {f - \nu } \right){e^{j2\pi f\tau }}{\rm{d}}f} , \label{C4_Ambiguity_function}
\end{align}
where ${X\left( f \right)}$ and ${Y\left( f \right)}$ are the corresponding Fourier transforms of $x\left( t \right)$ and $y\left( t \right)$, respectively.
Then, given ${{\cal Z}_x}\left( {\tau ,\nu } \right)$ and ${{\cal Z}_y}\left( {\tau ,\nu } \right)$ the Zak transforms of $x\left( t \right)$ and $y\left( t \right)$, we have
\begin{align}
{{\cal Z}_x}\left( {\tau ,\nu } \right)\! {\cal Z}_y^*\left( {\tau ,\nu } \right) \! =\!\!  \sum\limits_{n = \!  - \! \infty }^\infty  \! {\sum\limits_{m = \!  - \! \infty }^\infty \! \! \!  {{A_{x,y}}\! \! \left( {nT,\frac{m}{T}} \right)} } {e^{ - \! j2\pi n\nu T}}\! {e^{j2\pi \frac{m}{T}\tau }} .\label{C4_ZT_AF}
\end{align}
Conversely, we have
\begin{align}
{A_{x,y}}\! \left( {nT,\frac{m}{T}} \right) \! =\! \!  \int_0^T \! \! \! {\int_0^{\frac{1}{T}} \! \! \! {{{\cal Z}_x}\left( {\tau ,\nu } \right)\! \! {\cal Z}_y^*\left( {\tau ,\nu } \right)} } {e^{ - j2\pi \! \frac{m}{T}\! \tau }}\! {e^{j2\pi \!  n\nu T}} \! \! {\rm{d}}\nu {\rm{d}}\tau.\label{C4_AF_ZT}
\end{align}

For DD domain signal processing, the so-called ``fundamental rectangle'' is
the DD domain region of all $\left(\tau, \nu\right)$ defined by $\tau  \in \left[ {0,T} \right)$ and $\nu  \in \left[ {0,\frac{1}{T}} \right)$.
As suggested in Lemma~1, the Zak transform is uniquely defined by its response in the fundamental rectangle~\cite{Bolcskei1994Gabor}. 

\section{Communications with Delay-Doppler Signaling}

\begin{figure}
  \centering
  \includegraphics[width=0.4\textwidth]{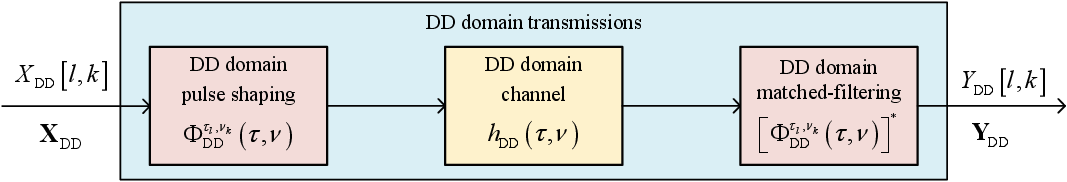}
  \caption{The considered DD domain transmissions for communications.}
  \label{System_model}
  \vspace{-3mm}
  \centering
  \end{figure}

We consider a type of DD signal transmissions for communications as shown in Fig.~\ref{System_model}, where
a set of discrete DD domain symbols are linearly modulated onto a family of continuous DD domain waveforms, i.e., \emph{DD domain basis functions}.
Let ${\bf X}_{\rm DD}$ be the DD domain symbol matrix of size $M \times N$, where $M$ and $N$ are numbers of delay and Doppler bins, respectively. Let ${X}_{\rm DD}\left[l,k\right]$ be the $\left(l,k\right)$-th element of ${\bf X}_{\rm DD}$, which is modulated onto the corresponding DD domain basis function ${\Phi _{{\rm{DD}}}^{{\tau _l},{\nu _k}}\left( {\tau ,\nu } \right)}$. Specifically, the considered linearly modulated DD  communication signal is formulated by
\begin{align}
{s_{{\rm{DD}}}}\left( {\tau ,\nu } \right) = \sum\limits_{l = 0 }^{M-1}  {\sum\limits_{k = 0 }^{N-1}  {{X_{{\rm{DD}}}}\left[ {l,k} \right]\Phi _{{\rm{DD}}}^{{\tau _l},{\nu _k}}\left( {\tau ,\nu } \right)} } ,\label{DD_modulated_signal}
\end{align}
where ${\Phi _{{\rm{DD}}}^{{\tau _l},{\nu _k}}\left( {\tau ,\nu } \right)}$ is the DD domain basis function with offset $\tau_l$ and $\nu_k$ and it will be detailed in the coming subsection.

As our main focus in this paper is the pulse shaping, we will not discuss the impact of the channel ${h_{{\rm{DD}}}}\left( {\tau ,\nu } \right)$ to DD communications. However, we highlight that the DD domain channel interacts with the DD domain signal in the form of \emph{twisted-convolution}, which essentially is a linear convolution with an additional phase term~\cite{MohammedBITSpart1,Mohammed2023part2}.
Let $r_{{\rm{DD}}}\left( \tau, \nu \right)$ be the received DD domain signal. Then, we have
\begin{align}
{Y_{{\rm{DD}}}}\left[ {l,k} \right] = \int_0^T {\int_0^{\frac{1}{T}} {{r_{{\rm{DD}}}}\left( {\tau ,\nu } \right)} } {\left[ {\Phi _{{\rm{DD}}}^{{\tau _l},{\nu _k}}\left( {\tau ,\nu } \right)} \right]^*}{\rm{d}}\nu {\rm{d}}\tau,
\label{DD_received_signal}
\end{align}
which is the sufficient statistics used for symbol detection.
\subsection{Delay-Doppler Domain Basis Functions}
Recall that $M$ and $N$ are the numbers of delay and Doppler bins within the fundamental rectangle, respectively.
Thus, a family of equally-spaced DD domain basis functions can be defined by
\begin{align}
{{\bm \Xi} _{{\rm{DD}}}}  \buildrel \Delta \over = & \left\{ \left. {\Phi _{{\rm{DD}}}^{{\tau _l},{\nu _k}}\left( {\tau ,\nu } \right)} \right|{\tau _l} = l \frac{T}{M},{v_k} = k \frac{1}{NT},\right.\notag\\
&\quad\left.l = \left\{ {0,...,M - 1} \right\},k = \left\{ {0,...,N - 1} \right\} \right\}. \label{DD_function_set}
\end{align}
Each element in ${\bm \Xi _{{\rm{DD}}}}$ is referred to as a DD domain basis function with delay and Doppler offsets $\tau_l$ and $\nu_k$.
In particular, we require that
\begin{align}
\Phi _{{\rm{DD}}}^{{\tau _l},{\nu _k}}\left( {\tau ,\nu } \right) = {e^{j2\pi {\nu_k}\left( {\tau  - {\tau _l}} \right)}}\Phi _{{\rm{DD}}}^{0,0}\left( {\tau  - {\tau _l},\nu  - {\nu_k}} \right),\label{DD_basis_offset}
\end{align}
holds for any $\tau\in \left( { - \infty ,\infty } \right)$ and $\nu\in \left( { - \infty ,\infty } \right)$. The phase term ${e^{j2\pi {\nu _k}\left( {\tau  - {\tau _l}} \right)}}$ in~\eqref{DD_basis_offset} comes from Lemma~2, and it is used to ensure that shifting the DD pulse $\Phi _{{\rm{DD}}}^{0,0}\left( {\tau ,\nu } \right)$ along delay and Doppler axes by ${{\tau _l}}$ and ${{\nu _k}}$ are corresponding to the application of time delay ${{\tau _l}}$ and phase rotations ${e^{j2\pi {\nu_k}\left( {t  - {\tau _l}} \right)}}$ to the time domain equivalent pulse{\footnote{Here, we assume that the delay operation goes first and the phase rotation goes second.}}, i.e., the time domain basis function, $\Phi _{{\rm{T}}}^{0,0}\left( t\right)$.
More specifically, applying the inverse Zak transform~\cite{Bolcskei1994Gabor} to $\Phi _{{\rm{DD}}}^{\tau_l,\nu_k}\left( \tau, \nu \right)$, the time domain basis function $\Phi _{{\rm{T}}}^{\tau_l,\nu_k}\left( \tau, \nu \right)$ can be shown to satisfy
\begin{align}
\Phi _{\rm{T}}^{{\tau _l},{\nu _k}}\left( t \right) \!=\! \sqrt T \!\int_0^{\frac{1}{T}} \!\!{\Phi _{{\rm{DD}}}^{{\tau _l},{\nu _k}}\left( {t,\nu } \right){\rm{d}}\nu }  = {e^{j2\pi {\nu_k}\left( {t - {\tau _l}} \right)}}\Phi _{\rm{T}}^{0,0}\left( {t - {\tau _l}} \right).\label{time_basis_offset_property}
\end{align}
In what follows, we shall refer to~\eqref{DD_basis_offset} as the
\emph{TF-consistency condition}, and we say a family of DD domain basis functions are  \emph{TF-consistent} if the functions within ${{\bm \Xi} _{{\rm{DD}}}}$ satisfy~\eqref{DD_basis_offset}. In particular, for TF-consistent family of DD domain basis functions, the following theorem holds.

\textbf{Theorem~1} (\emph{TF-consistency vs. Ambiguity Function}):
Let $\left(\tau_1, \nu_1\right)$, and $\left(\tau_2, \nu_2\right)$ be two pairs of arbitrary delay and Doppler offsets. Then, the following holds
\begin{align}
&\int_0^T {\int_0^{\frac{1}{T}} {{e^{j2\pi {\nu _2}\left( {\tau  - {\tau _2}} \right)}}{{\cal Z}_x}\left( {\tau  - {\tau _2},\nu  - {\nu _2}} \right)} }  \notag\\
&\quad{e^{ - j2\pi {\nu _1}\left( {\tau  - {\tau _1}} \right)}}{\cal Z}_x^*\left( {\tau  - {\tau _1},\nu  - {\nu _1}} \right){\rm{d}}\nu {\rm{d}}\tau\notag\\
=& {e^{j2\pi {\nu _2}\left( {{\tau _1} - {\tau _2}} \right)}}{A_x}\left( {{\tau _1} - {\tau _2},{\nu _1} - {\nu _2}} \right),
\label{DD_basis_function_00_vs_AF}
\end{align}
where ${A_{x}}\left( {{\Delta \tau},{\Delta \nu}} \right)$ denotes the self-ambiguity function of $x\left( {t } \right)$ with respect to the delay offset ${\Delta \tau}$ and Doppler offset ${\Delta \nu}$.
Particularly, for ${\Phi _{{\rm{DD}}}^{{\tau _1},{\nu _1}}\left( {\tau ,\nu } \right)}$ and ${\Phi _{{\rm{DD}}}^{{\tau _2},{\nu _2}}\left( {\tau ,\nu } \right)}$ belonging to ${{\bm \Xi} _{{\rm{DD}}}}$,~\eqref{DD_basis_function_00_vs_AF} suggests
\begin{align}
&\int_0^T {\int_0^{\frac{1}{T}} {\Phi _{{\rm{DD}}}^{{\tau _2},{\nu _2}}\left( {\tau ,\nu } \right){{\left( {\Phi _{{\rm{DD}}}^{{\tau _1},{\nu _1}}\left( {\tau ,\nu } \right)} \right)}^*}{\rm{d}}\nu {\rm{d}}\tau } }\notag\\
=&{e^{j2\pi {\nu _2}\left( {{\tau _1} - {\tau _2}} \right)}}{A_\Phi }\left( {{\tau _1} - {\tau _2},{\nu _1} - {\nu _2}} \right),
\label{DD_MF_basis_functions}
\end{align}
where ${A_{\Phi}}\left( {{\Delta \tau},{\Delta \nu}} \right)$ denotes the self-ambiguity function of $\Phi _{\rm{T}}^{0,0}\left( {t } \right)$.

\textbf{Proof}: The theorem can be straightforwardly derived based on Lemma~2 and Lemma~3. \hfill $\blacksquare$

Notice that~\eqref{DD_MF_basis_functions} is the DD domain matched-filtering.
Essentially, the property in~\eqref{DD_MF_basis_functions} suggests that the DD domain signal transmission with a family of TF-consistent DD domain basis functions can be characterized the ambiguity function of $\Phi _{\rm{T}}^{0,0}\left( {t } \right)$. Given above, we shall view ${\Phi _{{\rm{DD}}}^{{0},{0}}\left( {\tau ,\nu } \right)}$ as the DD domain \emph{prototype pulse} for DD pulse shaping.

\subsection{Constructing Delay-Doppler Domain Basis Functions}
Note that any DD domain signal satisfying the quasi-periodicity property can be sufficiently characterized by the corresponding pulse in the fundamental rectangle. Thus, we are motivated to construct the DD domain basis function by extending the ``atom pulse'' in the fundamental rectangle.
Let $\varphi \left( {\tau ,\nu } \right)$ be the atom pulse, whose support is the fundamental rectangle. According to Lemma~1, ${\Phi _{{\rm{DD}}}^{0,0}\left( {\tau ,\nu } \right)}$ can then be obtained by quasi-periodically extending $\varphi \left( {\tau ,\nu } \right)$, such as

\begin{align}
\Phi _{{\rm{DD}}}^{0,0}\left( {\tau ,\nu } \right) \buildrel \Delta \over = \sum\limits_{n \!= \! -\! \!\infty }^\infty  {\sum\limits_{m \!=\!  -\! \!\infty }^\infty  {\varphi \left( {\tau  - nT,\nu  - \frac{m}{T}} \right)} } {e^{j2\pi n\nu T}}.\label{DD_basis_function_construction}
\end{align}

\noindent By substituting~\eqref{DD_basis_function_construction} into~\eqref{DD_basis_offset}, we obtain
\begin{align}
&\Phi _{{\rm{DD}}}^{{\tau _l},{\nu _k}}\left( {\tau ,\nu } \right) \notag\\
=&\! \!\!\sum\limits_{n =  - \infty }^\infty \! {\sum\limits_{m =  - \infty }^\infty \! \!\!\!{{e^{j2\pi {\nu _k}\left( {\tau  \!-\! {\tau _l}} \right)}}\varphi\! \left( {\tau \! -\! {\tau _l} \!- \!nT,\nu \! - \!{\nu _k} \!- \!\frac{m}{T}} \right)} } \!{e^{j2\pi n\left( {\nu \! - \!{\nu _k}} \right)T}}.\label{DD_basis_function_offset_construction}
\end{align}

\begin{figure}
  \centering
  \includegraphics[width=0.4\textwidth]{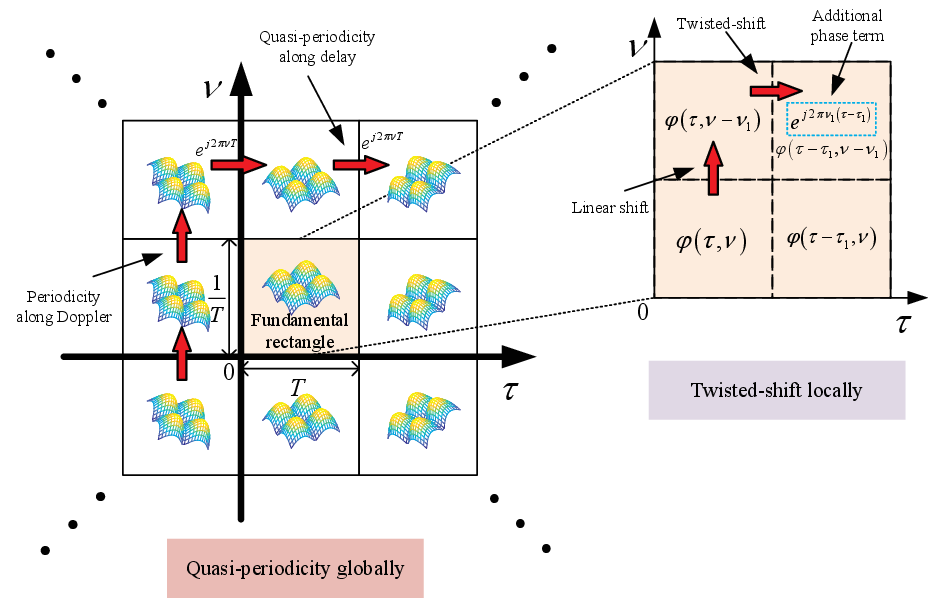}
  \caption{DD domain basis functions exhibit quasi-periodicity globally, which is constructed by locally twisted-shifting the atom pulse $\varphi \left( {\tau ,\nu } \right)$. Here, we assume $M=N=2$.}
  \label{DD_basis_fig}
  \vspace{-5mm}
  \centering
  \end{figure}

\noindent  We summarize the properties of the DD domain basis functions in Fig.~\ref{DD_basis_fig}, where we assume that $M=N=2$ such that there are $MN=4$ DD domain basis functions in the fundamental rectangle. We observe that the DD domain basis functions exhibit quasi-periodicity globally, and its constructed by locally twisted-shifting the atom pulse $\varphi \left( {\tau ,\nu } \right)$.
Specifically, the quasi-periodicity aligns with the property of the Zak transform, which characterizes the global structure of the DD domain basis functions across infinite numbers of regions of the size of the fundamental rectangle. On the other hand, the twisted-shift aligns with the property of the twisted convolution~\cite{MohammedBITSpart1,Mohammed2023part2}, characterizing the local structure of the DD domain basis functions across $M$ delay bins and $N$ Doppler bins.
Particularly, the global and local characteristics allow the DD domain basis functions have direct time domain and frequency domain interpretations via exploiting the properties of the Zak transform. Specifically, we can obtain the time domain and frequency domain basis functions by

%
%


  {\footnotesize\begin{align}
  &\Phi _{\rm{T}}^{{\tau _l},{\nu _k}}\left( t \right) ={e^{j2\pi {\nu_k}\left( {t - {\tau _l}} \right)}}\Phi _{\rm{T}}^{0,0}\left( {t - {\tau _l}} \right)\notag\\
=& \sqrt T {e^{j2\pi {\nu_k}\left( {t - {\tau _l}} \right)}}\sum\limits_{n =  - \infty }^\infty  {\int_{-\infty}^\infty {\varphi \left( {t - {\tau _l}- nT,\nu } \right){e^{j2\pi n\nu T}}{\rm{d}}\nu } } ,
  \label{time_basis_function_lk}
  \end{align}}
  and
  {\footnotesize
  \begin{align}
  &\Phi _{\rm{F}}^{{\tau _l},{\nu _k}}\left( f \right) = {e^{ - j2\pi f{\tau _l}}}\Phi _{\rm{F}}^{0,0}\left( {f - {\nu _k}} \right) \notag\\
  =& \frac{1}{{\sqrt T }}{e^{ - j2\pi f{\tau _l}}}\sum\limits_{m =  - \infty }^\infty  {\int_{-\infty}^\infty {\varphi \left( {\tau ,f - {\nu _k}- \frac{m}{T}} \right){e^{ - j2\pi f\tau }}{\rm{d}}\tau } } ,
  \label{frequency_basis_function_lk}
  \end{align}}

  \noindent where the detailed derivations are omitted due to the space limitation.
  Based on~\eqref{time_basis_function_lk} and~\eqref{frequency_basis_function_lk}, we observe that
  the DD domain basis functions can be understood as a mixture of the time domain and frequency domain pulses/tones~\cite{MohammedBITSpart1,Mohammed2023part2}. Specifically, we have the following observations:
  \begin{itemize}
  \item \textbf{DD domain global properties characterize the time and frequency periodicity}: The DD domain global characteristics are translated into the summation and integral terms in~\eqref{time_basis_function_lk} and~\eqref{frequency_basis_function_lk}, which leads to a train of pulses in the time and frequency domains that are apart in time by $T$ and apart in frequency by $\frac{1}{T}$, respectively.
  \item \textbf{DD domain local properties characterize time and frequency tones}: The DD domain local characteristics are translated into the phase terms (signal tones) ${e^{j2\pi {\nu_k}\left( {t - {\tau _l}} \right)}}$ and ${e^{ - j2\pi f{\tau _l}}}$ in~\eqref{time_basis_function_lk} and~\eqref{frequency_basis_function_lk}.
  \item \textbf{Time and frequency pulsones}: The special signal structure of~\eqref{time_basis_function_lk} and~\eqref{frequency_basis_function_lk} is known as the \emph{pulsone}~\cite{MohammedBITSpart1,Mohammed2023part2}, which is essentially a \emph{pulse train} modulated by a \emph{complex tone}.
  \item \textbf{Time and frequency limiting cases}: By letting $T \to \infty$, the time domain basis functions are separated only by the time offset $\tau_l$, yielding a pure time division multiplexing (TDM)-type of signaling; By letting $T \to 0$, the frequency domain basis functions are separated only by the frequency offset $\nu_k$, yielding a pure frequency division multiplexing (FDM)-type of signaling.
  \end{itemize}

\begin{figure}
  \centering
  \includegraphics[width=0.4\textwidth]{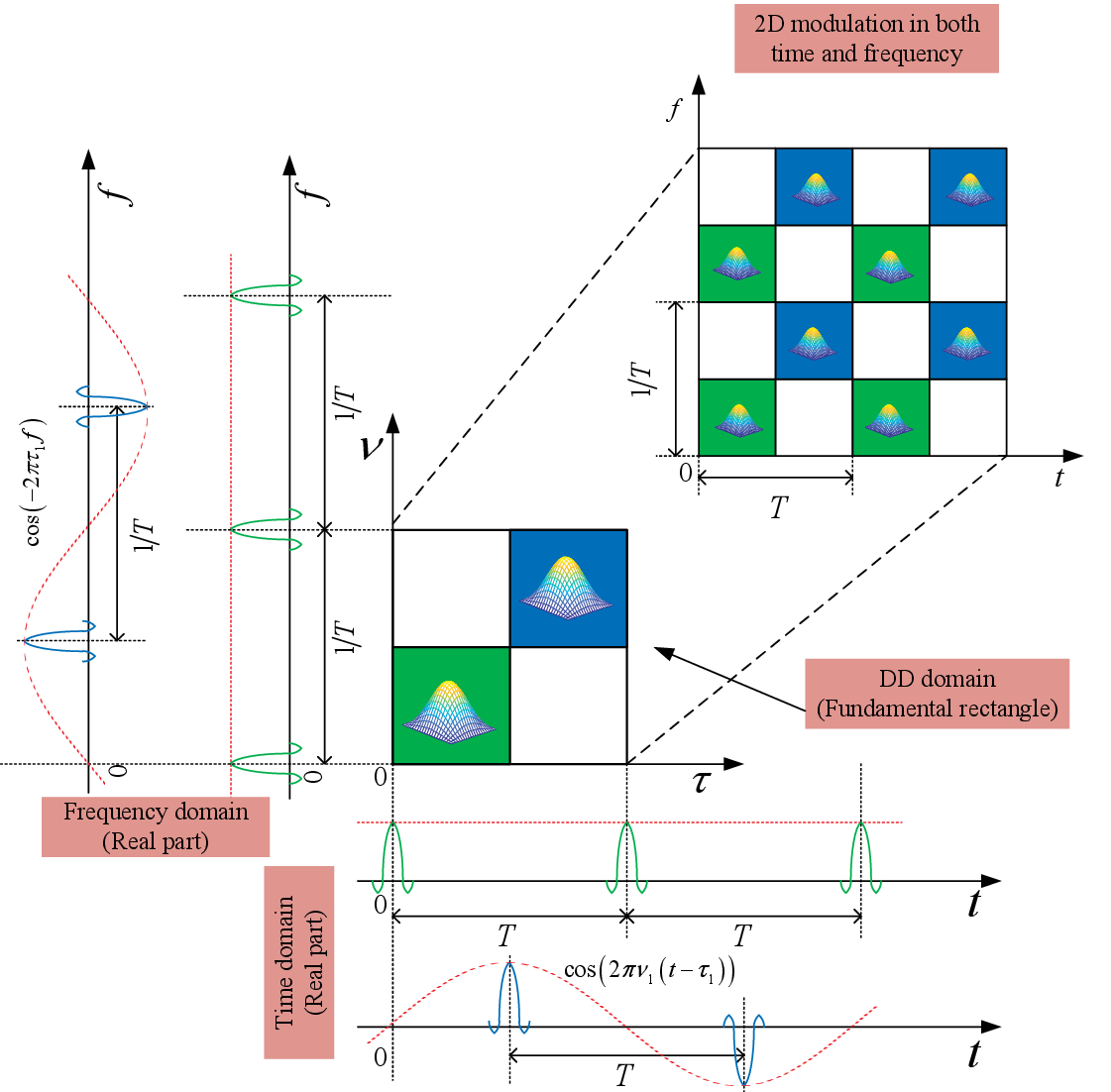}
  \caption{DD domain basis functions and equivalent representations in time and frequency domains. Here, the time and frequency pulses are shown with the same colors as the corresponding DD grids. In the figure, we assume $M=N=2$.}
  \label{DD_basis_T_F_fig}
  \vspace{-7mm}
  \centering
  \end{figure}

  We demonstrate the basis functions in different domains in Fig.~\ref{DD_basis_T_F_fig}, where we assume $M=N=2$ and mark the corresponding time and frequency pulses the same colors as those in the DD grids. As shown in the figure, the DD domain basis function becomes 2D pulsones in both time and frequency, while showing a particular response pattern according to the delay and Doppler offsets. Consequently, it is noticed that any symbol carried by the DD domain basis function will simultaneously be spread in both time and frequency in a periodic manner with respect to $T$ and $\frac{1}{T}$, leading to a potential of achieving full channel diversity{\footnote{We highlight here that the combination of time domain and frequency domain is not the commonly known time-frequency domain, where conventional OFDM multiplexes the information symbol.}}.

\section{Delay-Doppler Nyquist Pulse Shaping}
In this section, we study the ambiguity function and pulse shaping using the constructed basis functions. Due to the space limitation, we omit the detailed derivations and refer to the interested readers to the journal version of this paper.
We argue that the ideal DD domain basis functions may be a set of quasi-periodically extended delta functions, i.e., $\varphi \left( {\tau ,\nu } \right) = \delta \left( \tau  \right)\delta \left( \nu  \right)$, which are fully localized in the fundamental rectangle and therefore minimizes the interference among information symbols. It should be noted that the well-known \emph{Heisenberg's uncertainty principle} forbids the fully localized pulses in the TF domain~\cite{mallat1999wavelet}. However, the Heisenberg's uncertainty principle does not apply to the DD domain directly because delay and Doppler are not a pair of Fourier duals. More specifically, we shall highlight that the DD domain basis functions are only fully localized in the fundamental rectangle locally, while are in fact quasi-periodic in the whole DD domain globally according to~\eqref{DD_basis_function_offset_construction}.
Particularly, according to the definition of the Zak transform, a signal that is periodic in both time and frequency can yield

{\footnotesize\begin{align}
&\Phi _{{\rm{DD}}}^{{\tau _l},{\nu _k}}\left( {\tau ,\nu } \right) \notag\\
=& \!\sum\limits_{n =  - \infty }^\infty \! {\sum\limits_{m =  - \infty }^\infty \!\! {{e^{j2\pi {\nu _k}\left( {\tau \! -\! {\tau _l}} \right)}}\delta \left( {\tau  \!-\! {\tau _l} - nT} \right)\!\delta \left( {\nu \! -\! {\nu _k}\! - \!\frac{m}{T}} \right)} } {e^{j2\pi n\left( {\nu \! -\! {\nu _k}} \right)T}},
\label{DD_basis_localized}
\end{align}}
\noindent  whose ambiguity function satisfies
{\footnotesize
\begin{align}
{A_\Phi }\left( {{\tau _1} - {\tau _2},{\nu _1} - {\nu _2}} \right)
= \sum\limits_{n =  - \infty }^\infty  {\sum\limits_{m =  - \infty }^\infty  \delta  } \left( {{\tau _1}\! -\! {\tau _2} \!-\! nT} \right)\delta \left( {{\nu _1}\! -\! {\nu _2} \!-\! \frac{m}{T}} \right).
  \label{AF_fully_localized}
\end{align}}

\noindent The ambiguity function of the form of~\eqref{AF_fully_localized} minimizes the potential interference among different information symbols, but it only exists in theory, because periodic signals require infinite time duration or bandwidth theoretically. Therefore, we propose to apply practical filters and windows to limit the occupied TF resources.

\subsection{Time-Frequency Consistent Filtering and Windowing}
The idea of applying filtering and windowing for limiting the TF resources is straightforward, and previous implementation on OTFS based on this appears in~\cite{mohammed2021derivation}. However, what is not obvious and easy to be overlooked is the TF-consistency condition. We have shown in~\eqref{DD_basis_function_00_vs_AF} that the TF-consistency condition directly connects the DD domain matched-filtering and the ambiguity function. Consequently, filtering or windowing that does not align with the TF-consistency will break the DD domain integrity, and therefore degrades the communication performance.

Without loss of generality, we henceforth focus on the time domain TF-consistent filtering/windowing for the sake of practical implementation.
For a family of DD domain basis functions, we shall define the TF-consistent filtering/windowing in the following Proposition.

\textbf{Proposition~1} (\emph{TF-Consistent Filtering/Windowing}):
Define a family of DD domain basis functions ${{\bf{\Xi }}_{{\rm{DD}}}}$ that are delay and Doppler shifted with respect to a prototype pulse $\Phi _{{\rm{DD}}}^{0,0}\left( {\tau ,\nu } \right)$ in a TF-consistent manner, i.e.,~\eqref{DD_basis_offset}.
Define another family of DD domain basis functions ${{\bf{\tilde \Xi }}_{{\rm{DD}}}}$ that are obtained by time domain filtering or windowing each corresponding time domain basis function from
${{\bf{\Xi }}_{{\rm{DD}}}}$. We call the filtering/windowing is TF-consistent if and only if  ${{ \bf{\tilde \Xi }}_{{\rm{DD}}}}$ is TF-consistent.

Due to the space limitation, we cannot provide the detailed derivation for TF-consistent filtering/windowing. However, it can be shown that filtering/windowing a time domain signal in the following manner preserves the TF-consistency, i.e., $\tilde \Phi _{\rm{T}}^{{\tau _0},{\nu _0}}\left( t \right) = \Phi _{\rm{T}}^{{\tau _0},{\nu _0}}\left( t \right) \otimes \left( {{e^{j2\pi {\nu _0}t}}x\left( t \right)} \right)$ and $\tilde \Phi _{\rm{T}}^{{\tau _0},{\nu _0}}\left( t \right) = \Phi _{\rm{T}}^{{\tau _0},{\nu _0}}\left( t \right) \otimes \left( {{e^{j2\pi {\nu _0}t}}x\left( t \right)} \right)$.

According to Theorem~1, we notice that the ambiguity function of any function within the
family of TF-consistent DD domain basis functions can be characterized by the self-ambiguity function of the prototype pulse. Therefore, we will discuss the important properties of the self-ambiguity function of the prototype pulse in the following.

\subsection{Ambiguity Function of Truncated Basis Functions}
We restrict ourselves to only consider the case that the DD domain basis function is firstly truncated in frequency and then truncated in time following the TF-consistency condition. As a result, the equivalent time domain representation of the DD prototype pulse is given by $\tilde \Phi _{\rm{T}}^{0,0}\left( t \right) \buildrel \Delta \over = \left\{ {\Phi _{\rm{T}}^{0,0}\left( t \right) \otimes {\rm{F}}{{\rm{W}}_{\rm{T}}}\left( t \right)} \right\}{\rm{T}}{{\rm{W}}_{\rm{T}}}\left( t \right)$,
where ${{\rm{F}}{{\rm{W}}_{\rm{T}}}\left( t \right)}$ and ${{\rm{T}}{{\rm{W}}_{\rm{T}}}\left( t \right)}$ are the time domain representations of arbitrary frequency domain and time domain windows, respectively.
This gives rise to the following theorem.

\textbf{Theorem~2} (\emph{Time Frequency Truncation and Orthogonality}):
Consider a basis function constructed by delay and Doppler atom pulses with periodic duals, which is first frequency truncated and then time truncated. Its ambiguity function satisfies

{\footnotesize\begin{align}
{A_{\tilde \Phi }}\left( {\tau ,\nu } \right)\!=\!\sum\limits_{n =  - \infty }^\infty  {\sum\limits_{m =  - \infty }^\infty  {{A_{{\rm{FW}}}}} \left( {\tau \! - \! nT,\frac{m}{T}} \right)} {A_{{\rm{TW}}}}\left( {\tau ,\nu \! -\! \frac{m}{T}} \right).
\label{AF_trunction}
\end{align}}

\noindent  Theorem~2 states that the ambiguity function of truncated DD domain basis functions can be represented by infinite summations of the ambiguity functions of the adopted windows in both time and frequency.
From Theorem~2, various truncated DD domain basis functions with desired ambiguity function can be designed by carefully selecting the windows in time and frequency.
To further discuss the insights, let us consider time and frequency windows with specific constraints.
We consider the time domain window has a finite time duration from $t \in \left[ {0,\tilde NT} \right]$, while the frequency domain window has a finite bandwidth $f \in \left[ {0,\tilde M/T} \right]$, respectively, where $\tilde N \ge N$ and $\tilde M \ge M$. Furthermore, notice that the Zak transform involves periodic summations. We are motivated to consider \emph{periodic windows} in time and frequency for basis function truncation.
We call a time domain window ${{\rm{T}}{{\rm{W}}_{\rm{T}}}\left( t \right)}$ $T$-periodic if ${{\rm{T}}{{\rm{W}}_{\rm{T}}}\left( t \right)}={{\rm{T}}{{\rm{W}}_{\rm{T}}}\left( t+T \right)}$, for $t \in \left[ {0,\tilde NT} \right]$. Similarly, a frequency domain $\tilde M/T$-periodic window satisfies ${{\rm{F}}{{\rm{W}}_{\rm{F}}}\left( f \right)}={{\rm{F}}{{\rm{W}}_{\rm{F}}}\left( f+1/T \right)}$, for $f \in \left[ {0,\tilde M/T} \right]$.
With time and frequency periodic windows described above, we have the following corollary.

\textbf{Corollary~1} (\emph{Ambiguity Function of Truncated Basis Function with Periodic Windows}):
For DD domain basis functions of the form~\eqref{DD_basis_localized} that are frequency and time truncated by energy-normalized periodic windows, under
sufficiently large $\tilde M$ and $\tilde N$, its ambiguity function can be approximated by 

{\footnotesize \begin{align}
{A_{\tilde \Phi }}\left( {\tau_1 ,\nu_1 } \right)\! \approx \!\tilde M\tilde N\frac{{{{\left| {{\rm{F}}{{\rm{W}}_{\rm{F}}}\left( 0 \right)} \right|}^2}}}{T}\!\!\int_0^T \!\!{{\rm{T}}{{\rm{W}}_{\rm{T}}}\left( \tau  \right){\rm{TW}}_{\rm{T}}^*\left( {\tau  - {\tau _1}} \right)} {\rm{d}}\tau \delta \left[ {{l_1}} \right]\delta \left[ {{k_1}} \right], \label{AF_trunction_periodic_bound}
\end{align}}

\noindent where ${\tau _1} = {l_1}\frac{T}{{\tilde M}}$ and ${\nu _1} = \frac{{{k_1}}}{{\tilde NT}}$ with $ - \tilde M \le {l_1} \le \tilde M$ and $ - \tilde N \le {k_1} \le \tilde N$. Furthermore, the approximation becomes exact with rectangular windows, where~\eqref{AF_trunction_periodic_bound} reduces to

{\footnotesize 
\begin{align}
{A_{\tilde \Phi }}\left( {\tau_1 ,\nu_1 } \right) ={\tilde M}{\tilde N}{\left| {{\rm{F}}{{\rm{W}}_{\rm{F}}}\left( 0 \right)} \right|^2}{\left| {{\rm{T}}{{\rm{W}}_{\rm{T}}}\left( 0 \right)} \right|^2}\delta \left[ {{l_1}} \right]\delta \left[ {{k_1}} \right],
\end{align}}

\noindent  In fact, the ambiguity function ${A_{\tilde \Phi }}\left( {\tau_1 ,\nu_1 } \right)$ is dominated by \emph{aliased sinc (asinc) functions}, which are a special type of quasi-orthogonal signals with respect to the interval $\frac{T}{\tilde M}$ and $\frac{1}{\tilde NT}$, except for time and frequency instants at integer times of $T$ and $\frac{1}{T}$, respectively.
Furthermore, Corollary~1 suggests that different periodic windows may share similar ambiguity function shapes that are dominated by the asinc functions, and as a result achieves the approximate DD orthogonalities.
This fact greatly relaxes the requirements of pulse designs for DD communications.
Notice that root-raised cosine (RRC) windows with small roll-off factors, i.e., a smoothed rectangular window with marginal excessive bandwidth/time duration, can be viewed as an approximated periodic windows.
Therefore, we may use either RRC windows or periodic windows interchangeably for achieving DD Nyquist signaling in practice. As a matter of fact, a DD domain signaling of using both RRC window and rectangular window appears in~\cite{Lin2022ODDM}.

\textbf{Remark}: In fact, the localization in~\eqref{AF_fully_localized} and the orthogonality suggested by~\eqref{AF_trunction_periodic_bound} are well-aligned with the intuitions of delay and Doppler by considering the time and frequency partition under critical sampling. Intuitively, the delay and Doppler implies how significant the signal changes in frequency and in time.
Clearly, for periodic signals in time and frequency, their Doppler and delay responses will be fully localized, as will their ambiguity functions. However, for periodic signals with truncation, their delay and Doppler response will not be fully localized, because the signal periodicity is broken due to the truncation.
Consequently, their Zak transforms will be sufficiently concentrated depending on the duration of the truncation window in time and frequency, following a ``sinc-like'' pattern. This is because the truncation in time and frequency can be viewed as the multiplication of time and frequency rectangular windows, and the ``sinc-like'' pattern appears naturally as the result of the Zak transform to a rectangular window.

\subsection{Delay-Doppler Nyquist Pulse Shaping}
\begin{figure}
  \centering
  \includegraphics[width=0.4\textwidth]{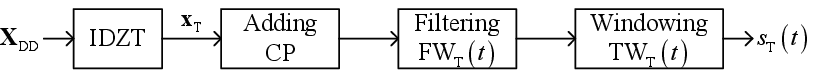}
  \caption{The diagram of the proposed DD Nyquist pulse shaping.}
  \label{DD_pulse_shaping}
  \vspace{-6mm}
  \centering
  \end{figure}
Following the previous discussions, we can further derive the time domain transmitted signal according to the TF-consistent filtering and windowing, such as

{\footnotesize \begin{align}
{s_{\rm{T}}}\!\left( t \right)
=& \sqrt {NT} \sum\limits_{l = 0}^{M - 1} \sum\limits_{n = 0}^{N - 1} {x_{\rm{T}}}\left[ {l + nM} \right]\notag\\
&\quad\quad \sum\limits_{m =  - \infty }^\infty  {{\rm{F}}{{\rm{W}}_{\rm{T}}}\left( {t\! - \!\frac{l}{M}T \!-\! nT \!-\! mNT} \right)}   {\rm{T}}{{\rm{W}}_{\rm{T}}}\left( t \!-\! \frac{l}{M}T\right)\notag\\
\approx& \!\sqrt {NT}\!\! \sum\limits_{l = 0}^{M - 1} \!{\sum\limits_{n = 0}^{N - 1}\!\! {{x_{\rm{T}}}\left[ {l\! +\! nM} \right]\!{\rm{F}}{{\rm{W}}_{\rm{T}}}\!\left( {t \!-\! \!\frac{l}{M}T \!-\! nT} \right)} } {\rm{T}}{{\rm{W}}_{\rm{T}}}\!\left( t \right),
\end{align}}

\noindent where ${\bf x}_{\rm T}$ of length-$MN$ is the time domain symbol vector obtained by applying
the inverse discrete Zak transform (IDZT) to ${\bf X}_{\rm DD}$, i.e., ${x_{\rm{T}}}\left[ {l + nM} \right] = \frac{1}{{\sqrt N }}\sum\nolimits_{k = 0}^{N - 1} {{x_{{\rm{DD}}}}\left[ {l,k} \right]} {e^{j2\pi k\frac{n}{N}}}$. Note that the term $\sqrt {NT}$ is roughly the time duration of the signal transmission, which ensures each time domain symbol is modulated by a unit-energy pulse.
As such, we propose the transmitter diagram in Fig.~\ref{DD_pulse_shaping} for DD Nyquist pulse shaping.
In Fig.~\ref{DD_pulse_shaping}, the DD domain symbol matrix ${\bf X}_{\rm DD}$ is first passed to the inverse discrete Zak transform (IDZT), which is essentially the discrete version of the inverse Zak transform, yielding ${\bf x}_{\rm T}$ of length-$MN$.
After appending a cyclic-prefix (CP) with duration longer than the maximum path delay{\footnote{The reason for appending such a CP is to ensure the quasi-periodicity of the received symbol sequences after passing over the channel.}}, the resultant vector is first convoluted with ${\rm{F}}{{\rm{W}}_{\rm{T}}}\left( t \right)$ and then windowed by ${\rm{T}}{{\rm{W}}_{\rm{T}}}\left( t \right)$. The underlying window functions can be selected according to our previous discussions.
\section{Numerical Results}
We examine the proposed DD Nyquist pulse shaping as follows. We consider $M=N=32$, and $T=1$, respectively. We consider three different windows, namely the rectangular window, RRC windows with roll-off factor $\beta=0.3$, and the truncated $T$-periodic cosine pulse, respectively. Specifically, we study the following three cases: 1) both time and frequency windows are rectangular windows with duration $NT$ and $\frac{M}{T}$ (termed as ``Sinc + Sinc''); 2) both time and frequency windows are RRC windows, with orthogonal periods being $\frac{1}{NT}$ and $\frac{T}{M}$, respectively (termed as ``RRC + RRC''); 3) the time window is a truncated cosine function with duration $NT$ and the frequency window is the RRC window with orthogonal period $\frac{M}{T}$ (termed as ``Cos + RRC'').

\begin{figure}
    \centering
    \includegraphics[width=0.4\textwidth]{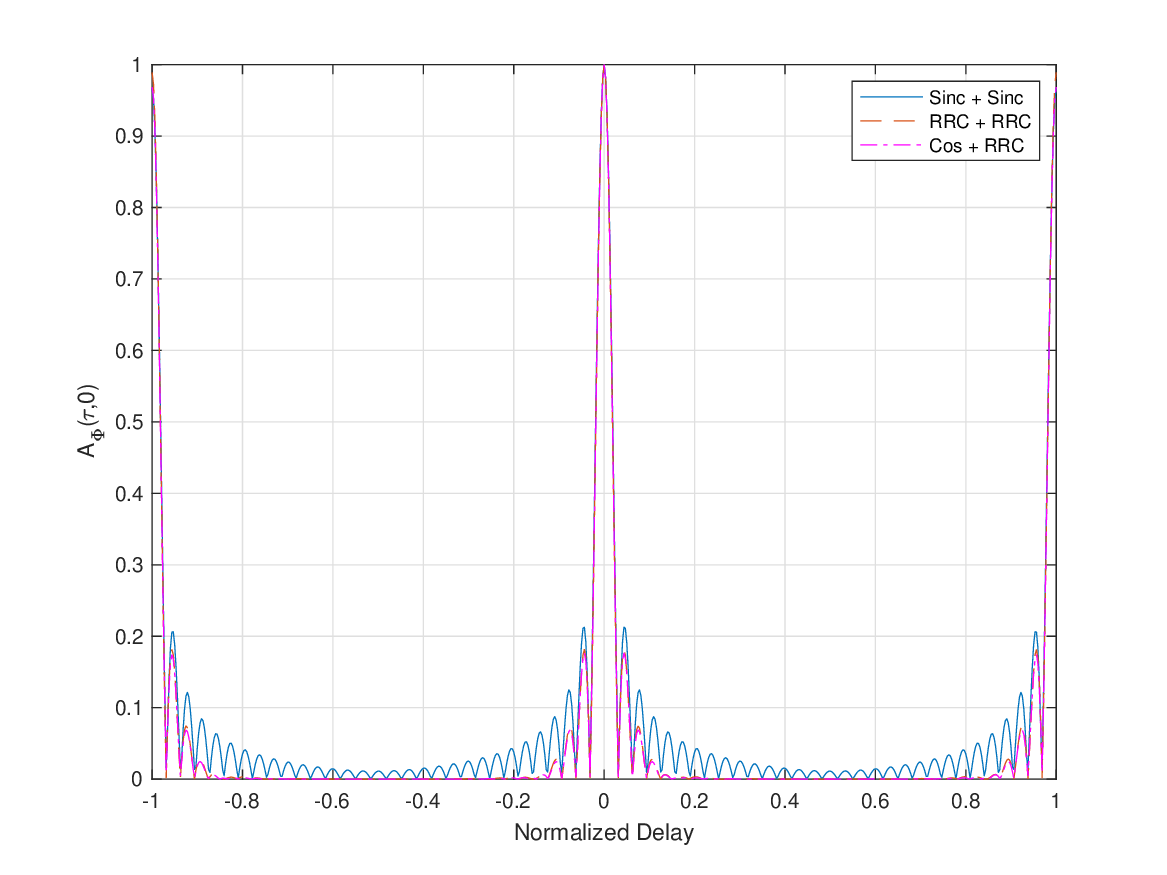}
    \caption{Zero-Doppler cut of the ambiguity function.}
    \label{zero_Doppler}
    \centering
    \vspace{-3mm}
    \end{figure}

    \begin{figure}
      \centering
      \includegraphics[width=0.4\textwidth]{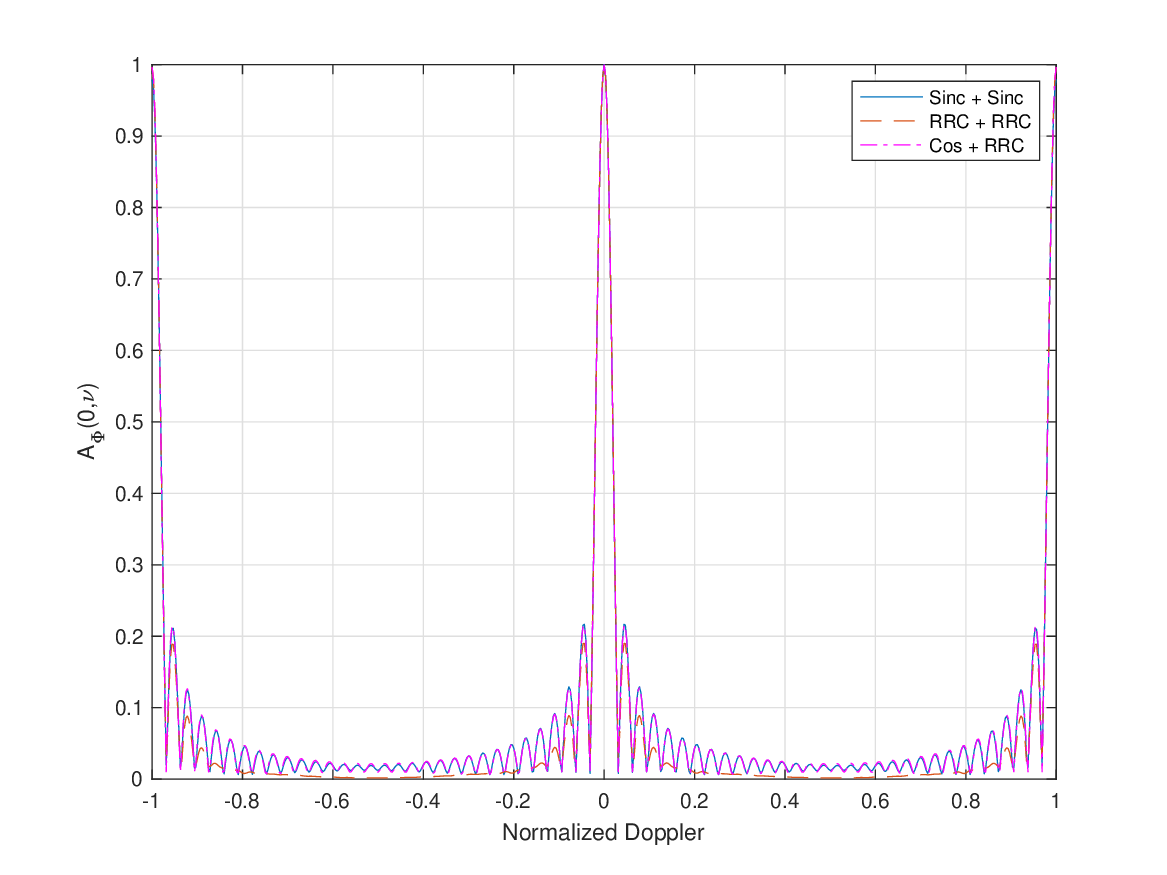}
      \caption{Zero-delay cut of the ambiguity function.}
      \label{zero_delay}
      \centering
      \vspace{-5mm}
      \end{figure}

We present the zero-Doppler cuts (ambiguity function with zero Doppler) of the three cases in Fig.~\ref{zero_Doppler}, where it is observed that all the three cases can achieve the sufficient delay orthogonality. Furthermore, we observe that both case-2 and case-3 have an almost zero response for normalized delay around $-\frac{T}{2}$ and $\frac{T}{2}$. This is due to the quick decay of RRC pulses at the cost of the excess bandwidth. As a result, we can observe a ``spike-like'' ambiguity function, which may also be of interest for radar sensing.

The performance of zero-delay cut (ambiguity function with zero delay) is presented in Fig.~\ref{zero_delay}. From the figure, we observe that all three cases can achieve sufficient Doppler orthogonality. Again, we observe that using RRC windows can achieve almost zero response at normalized Doppler around $-\frac{1}{2T}$ and $-\frac{1}{2T}$ thanks to its quick decay property. Furthermore, we notice that using truncated cosine window in time share almost the same zero-delay cut as using the rectangular window, which aligns with our discussion in Corollary~1, i.e., periodic windows essentially share very similar ambiguity functions for DD communications. However, it should be noted that using different periodic windows may result in different transmit signal spectrums, and we can design the window response in order to achieve the desired spectrum shape.

\section{Conclusions}
In this paper, we study the pulse shaping for DD communications. By constructing a DD domain basis function from atom pulses and studying its properties, we demonstrated some important observations on its ambiguity function. Based on those observations, we further proposed a DD Nyquist pulse shaping scheme. Our conclusions were also verified by numerical simulations.

\bibliographystyle{IEEEtran}
\bibliography{OTFS_references}

\end{document}